\documentclass[journal]{IEEEtran}
\usepackage{amsmath,amssymb,amsfonts}
\usepackage{xurl} 
\usepackage{graphicx}
\usepackage{subcaption}
\usepackage{url}
\usepackage{listings} 
\usepackage{xcolor}
\usepackage{tabularx}
\usepackage{booktabs}
\newcolumntype{L}{>{\RaggedRight\hangafter=1\hangindent=0em}X}

\definecolor{mygreen}{rgb}{0,0.6,0}
\definecolor{mymauve}{rgb}{0.58,0,0.82}
\begin{document}

\title{Subgraph-Oriented Testing for Deep Learning Libraries}

\author{
\IEEEauthorblockN{
Xiaoyuan Xie\IEEEauthorrefmark{1}\IEEEauthorrefmark{3}\thanks{\IEEEauthorrefmark{3}Xiaoyuan Xie and Songqiang Chen are the co-corresponding authors.},
Yan Song\IEEEauthorrefmark{1},
Songqiang Chen\IEEEauthorrefmark{2}\IEEEauthorrefmark{3},
Jinfu Chen\IEEEauthorrefmark{1}
}

\IEEEauthorblockA{\IEEEauthorrefmark{1}School of Computer Science, Wuhan University, China}

\IEEEauthorblockA{\IEEEauthorrefmark{2}Department of Computer Science and Engineering, The Hong Kong University of Science and Technology, China}

\IEEEauthorblockA{xxie@whu.edu.cn, yansong@whu.edu.cn, i9s.chen@connect.ust.hk, jinfuchen@whu.edu.cn}
}

\maketitle

\begin{abstract}
Deep Learning (DL) libraries, such as PyTorch, are widely used for building and deploying DL models on various hardware platforms. Meanwhile, they are found to contain bugs that lead to incorrect calculation results and cause issues like non-convergence training and inaccurate prediction of DL models. Thus, many efforts have been made to test DL libraries and reveal bugs. 
However, existing DL library testing methods manifest limitations: model-level testing methods cause complexity in fault localization. Meanwhile, API-level testing methods often generate invalid inputs or primarily focus on extreme inputs that lead to crash failures; they also ignore testing realistic API interactions. These limitations may lead to missing detection of bugs, even in the frequently used APIs.
To address these limitations, we propose \textbf{SORT} (\textbf{S}ubgraph-\textbf{O}riented \textbf{R}ealistic \textbf{T}esting) to differential test DL libraries on different hardware platforms. 
SORT takes popular API interaction patterns, represented as frequent subgraphs of model computation graphs, as test subjects. In this way, it introduces realistic API interaction sequences while maintaining efficiency in locating faulty APIs for observed errors. 
Besides, SORT prepares test inputs by referring to extensive features of runtime inputs for each API in executing real-life benchmark data. The generated inputs are expected to better simulate such valid real inputs and reveal bugs that are more likely to happen in real-life usage.
Evaluation on 728 frequent subgraphs of 49 popular PyTorch models demonstrates that SORT achieves a 100\% valid input generation rate, detects more precision bugs than existing methods, and reveals interaction-related bugs missed by single-API testing. 18 precision bugs in PyTorch are identified and reported to PyTorch developers.
\end{abstract}


\section{Introduction}
\label{section_1}

Deep Learning (DL) techniques have been widely adopted in various fields of our daily life, such as recommendation, visual perceptual recognition, and text-processing \cite{computational_2025_niu, DeepLearningRecommender_2021_Steck, DeepLearningClassification_2021_Minaee, Image_2021_Minaee, Prediction_2021_Bharti, LargeSequence_2023_Wen}. 
In building, training, and deploying DL models, DL libraries like PyTorch \cite{Pytorch_2019_Paszke} are widely adopted to perform mathematical computations across diverse hardware environments \cite{survey_2024_zhang}. 
However, like other software systems, DL libraries also contain bugs. Such bugs can lead to serious issues such as non-convergence training \cite{RealWorldNumericalBugs_2017_di} and inaccurate prediction \cite{CRADLE_ICSE_2019_Pham}. 
Therefore, testing DL libraries has attracted many studies, aiming to uncover hidden bugs and promote the repair for DL libraries.

To test DL libraries, practitioners generate test inputs for DL library APIs and validate the API outputs \cite{Generation_2023_liu}. Earlier DL library testing methods apply on the \textit{model-level} \cite{CRADLE_ICSE_2019_Pham, Audee_ASE_2020_Guo, LEMON_FSE_2020_Wang}. 
They directly execute an entire DL model to evaluate the functionality of the involved DL library APIs. While model-level DL library testing helps validate API behavior during model training or inference, it introduces complexity in fault localization. 
Numerous APIs invoked in one test execution make it difficult to identify the root cause of bugs~\cite{FreeFuzz_ICSE_2022_Wei}, which is a tedious and time-consuming process for developers \cite{improving_2024_zhang}.

To avoid the shortcomings of the bulky model-level testing, many \textit{API-level} DL library testing methods have emerged recently \cite{FreeFuzz_ICSE_2022_Wei, DocTer_ISSTA_2022_Xie, DeepREL_FSE_2022_Deng, TitanFuzz_ISSTA_2023_Deng, ACETest_2023_Shi, Predoo_ISSTA_2021_Zhang, NablaFuzz_ICSE_2023_Yang}. 
Such methods test APIs in a DL library individually. Specifically, they generate test inputs for one single API and check the correctness of corresponding outputs. 
Such testing methods address the limitations of model-level methods and have become a popular paradigm. However, they also manifest three significant problems. 

\textbf{Firstly, existing API-level testing methods \cite{FreeFuzz_ICSE_2022_Wei, DeepREL_FSE_2022_Deng, DocTer_ISSTA_2022_Xie, ACETest_2023_Shi, TitanFuzz_ISSTA_2023_Deng} often generate inputs unable to pass API input validity check} (e.g., due to incorrect tensor shapes). 
While some methods \cite{DocTer_ISSTA_2022_Xie, ACETest_2023_Shi} try to extract constraints to guide the generation of valid inputs, they still manifest a low ratio of valid inputs because of the incomplete documents and limited information able to be extracted from the code. 
\textbf{Secondly, even if some test inputs pass the validity check, many of them are boundary or extreme values that mainly lead to crash failures.} 
Meanwhile, precision bugs, another type of important fault that may significantly damage the reliability of high-performance computing and deep learning models \cite{EfficientFloatingPointInput_2020_guo}, are rarely discovered.

\textbf{Besides, traditional API-level methods fail to involve real interactions between APIs.} 
Interactions among APIs in sequential execution of multiple APIs may trigger failures. However, traditional API-level methods \cite{FreeFuzz_ICSE_2022_Wei,DeepREL_FSE_2022_Deng,DocTer_ISSTA_2022_Xie,ACETest_2023_Shi,NablaFuzz_ICSE_2023_Yang,Predoo_ISSTA_2021_Zhang} only invoke single APIs at a time and cannot test such interactions among APIs. 
While some approaches \cite{GraphBased_ICSE_2021_Luo,Muffin_ICSE_2022_Gu,COMET_TOSEM_2023_Li, TitanFuzz_ISSTA_2023_Deng,EAGLE_ICSE_2022_Wang} consider combining APIs artificially, such arbitrary combinations may not match the popular API interaction patterns in the real-life usage. The identified errors are thereby less likely to happen in real-world scenarios, and as a result, developing efforts may be distracted from the errors that are more likely to happen in real-life usage.

In this paper, we propose a novel DL library testing method called \textbf{SORT} to conduct \textbf{S}ubgraph-\textbf{O}riented \textbf{R}ealistic \textbf{T}esting to overcome the limitations of existing methods. 
Specifically, \textbf{to solve the lack of realistic interactions} while maintaining the efficiency in locating faulty APIs, SORT considers test subjects of a new granularity, i.e., frequent subgraphs of model computation graphs. 
Frequent subgraphs help to reflect popular API interaction patterns in real-life usage while including only a few APIs in a test subject. 
\textbf{To generate valid inputs} for test subject subgraphs, SORT refers to the runtime inputs of APIs during running models with real datasets. By generating inputs whose type, shape, and value range are similar to the runtime inputs of APIs, SORT tends to generate inputs that are more likely valid and appear in real-life usage.
   
Based on the above two ideas, SORT runs four steps. 
Firstly, it extracts frequent computation subgraphs from popular open-source DL models. 
Then, it instruments the code and runs DL models with realistic datasets to collect the features of runtime input data for each API in the collected models. 
Afterward, it generates test inputs for frequent subgraphs based on the collected information. Finally, it performs differential testing over different hardware (e.g., CPU and GPU) to reveal errors. 

On 728 subgraphs extracted from 49 popular DL models implemented in PyTorch, we performed differential testing across CPU and GPU using SORT to evaluate its effectiveness. 
Results show that SORT can effectively reveal precision errors with valid inputs. 
It shows a 100\% validity rate of generated inputs, outperforming all baselines. 
Furthermore, SORT detects more precision errors than baselines. It is also found that many reported bugs initially appear as small precision losses, which cannot be detected without considering API interactions. In addition, we conduct a survey to collect developers' thoughts about DL library testing. The feedback confirms the helpfulness of taking frequent graphs as testing subjects in introducing meaningful API interactions while maintaining fault localization efficiency, as well as the significance of the revealed precision bugs signified by API interactions.

The contributions of this paper are summarized as follows:
\begin{itemize}
    \item We propose to test DL libraries with a new granularity of test subjects, i.e., frequent subgraphs of model computation graphs. Testing with frequent subgraphs introduces more meaningful API interactions to expose the impactful output differences that are more likely to happen in the real-life usage of DL libraries while maintaining efficiency in locating the faulty APIs causing the observed differences. 
    \item We propose a method to generate test inputs by referring to the shape, data type, and value or value range features of the runtime API inputs captured by instrumentation. Such generated test inputs try to simulate the inputs in real-life model usage and thus are more likely to be valid.
    \item We evaluate the effectiveness of SORT and found that SORT can achieve a much higher (100\%) ratio of valid inputs and detect more precision errors than existing methods. We also found that SORT can reveal interaction-related errors that will be missed by single-API testing.
    \item  We identified 18 precision bugs of PyTorch.
\end{itemize}

We release the artifact of this paper including the replication package and additional materials online at \cite{artifact}.

\section{Motivation and Preliminary Study} 
\label{section_2}

\subsection{Limitations in Current Methods}
\label{section_2_1}

\subsubsection{\textbf{Lack of Valid Inputs}}
\label{section_2_1_1}

Valid inputs refer to the inputs that can pass APIs' input validity check (e.g., checking parameter type, tensor shape, etc.) and get a calculation result from APIs \cite{DocTer_ISSTA_2022_Xie, ACETest_2023_Shi}. 
The ratio of valid test inputs among all generated test inputs is expected to be high such that a testing method can produce many tests for validating the calculation function of APIs.

However, existing methods report a low valid rate of generated inputs. For example, DocTer \cite{DocTer_ISSTA_2022_Xie} and ACETest \cite{ACETest_2023_Shi} report valid rates of 12.38\% and 43.66\%{, respectively}.  
DocTer's low valid rate can be attributed to incomplete documentation \cite{DocTer_ISSTA_2022_Xie} and ACETest's relatively low ratio of valid inputs is due to limited information extracted from the code \cite{ACETest_2023_Shi}. 
We also reproduced the other popular existing testing methods that do not report the ratio of valid inputs, i.e., FreeFuzz \cite{FreeFuzz_ICSE_2022_Wei}, TitanFuzz \cite{TitanFuzz_ISSTA_2023_Deng}, and DeepREL \cite{DeepREL_FSE_2022_Deng}, and found that their average ratio of valid inputs is 16.01\%, 41.54\%, and 81.34\%, respectively. 
Overall, the ratio of valid inputs generated by current methods is unsatisfactory, mainly due to incomplete or unreliable information to guide generation. The detailed reasons will be discussed in Section \ref{section_5_1}. The unsatisfactory ratio of valid inputs may result in inadequate testing of APIs' calculation function \cite{ACETest_2023_Shi}.

\subsubsection{\textbf{Missing Reporting Precision Errors}}
\label{section_2_1_2}

\begin{figure}[htbp] 
\centering 
\includegraphics[width=1\linewidth]{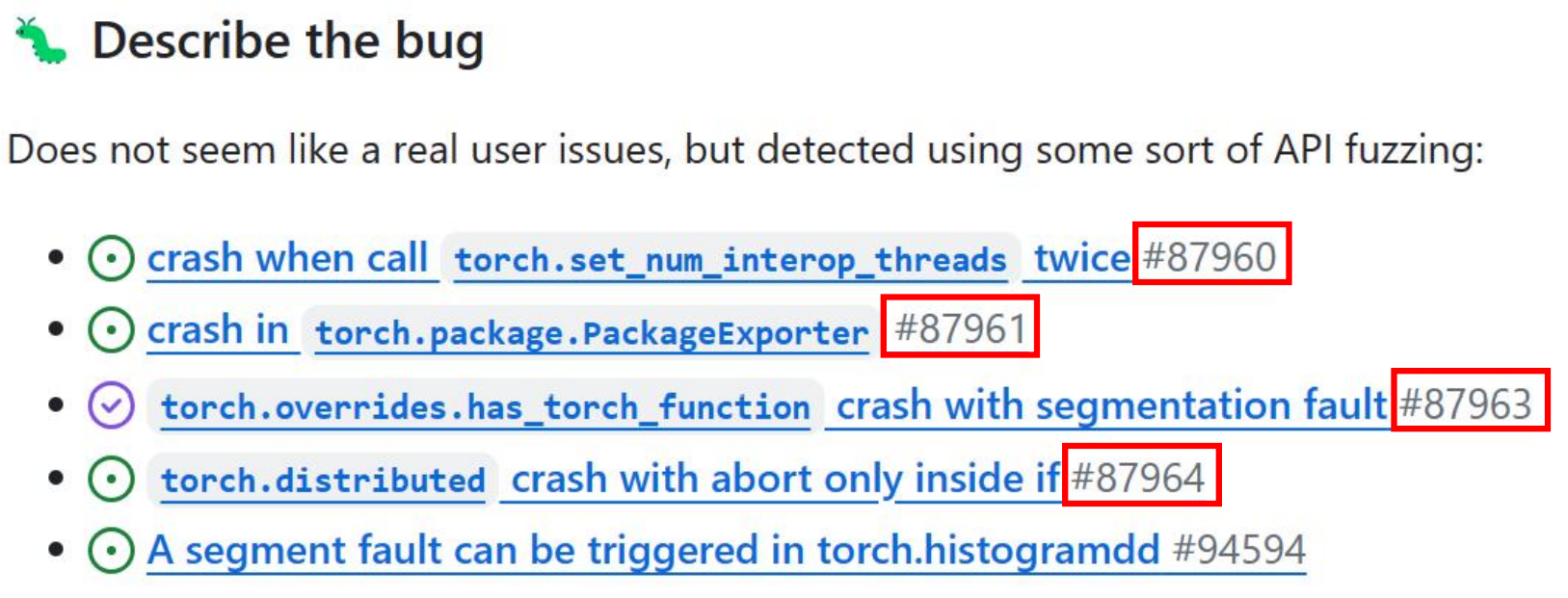} 
\caption{``Not Like Real User Issues'' Reported by TitanFuzz}
\label{motivation_not_real_user} 
\end{figure}

Even if existing methods can generate a few test inputs that pass APIs' validity check, most such inputs are corner cases that lead to crash failures, e.g., tensors with extreme values or unusual types. 
These crash failures tend to happen less frequently in real-life model deploying scenarios as suggested by developers. 
As shown in Fig.~\ref{motivation_not_real_user}, the PyTorch developers considered many crash issues reported by TitanFuzz (highlighted in a red box in the figure) as \textbf{\textit{``not real user issues''}}\footnote{https://github.com/pytorch/pytorch/issues/88148}. 
Meanwhile, other types of errors need to be exposed by non-corner cases. Precision errors are typical of such errors, which manifest as problem-inducing value differences (i.e., precision losses) between the correct calculation result and the actual results returned on certain computation hardware. They arise from the differences of unique hardware in handling limitations of finite precision and rounding errors inherent in floating-point representations \cite{Predoo_ISSTA_2021_Zhang, EfficientFloatingPointError_2014_chiang}. Such errors are considered a central concern in high-performance computing since significant precision errors will lead to inaccurate calculation results \cite{EfficientFloatingPointInput_2020_guo} and damage supercomputers and handheld electronics systems \cite{EfficientFloatingPointError_2014_chiang}.

However, current methods often overlook generating non-corner cases and seldom identify precision errors. 
For example, FreeFuzz, DeepREL, DocTer, TitanFuzz, and ACETest have identified a total of 143 issues of PyTorch; but only four of them are precision-related: FreeFuzz found one, TitanFuzz reported three, and DeepREL, DocTer, and ACETest reported no precision issues. 
Although Predoo and NablaFuzz aim to identify precision errors, they only apply to a few specific APIs. 
Predoo only applies to seven APIs with known susceptibility to precision issues and NablaFuzz only focuses on automatic differentiation components for gradient calculation.

\subsubsection{\textbf{Lack of Realistic API Interactions}}
\label{section_2_1_3}

Interactions among APIs may help to magnify a small difference in the output from the previous API. 
For example, Fig.~\ref{motivation_single_api} illustrates differential testing of two APIs (\texttt{layer\underline{~}norm} and \texttt{linear}) independently, and no precision differences between CPU and GPU reach the suggested thresholds, thereby no issue is detected in this case. 
Meanwhile, Fig.~\ref{motivation_interaction} shows an API interaction involving \texttt{layer\underline{~}norm} and \texttt{linear}. In this case, the output difference of \texttt{layer\underline{~}norm} did not reach the suggested thresholds, but the difference was magnified by the subsequent API \texttt{linear}. 
This interaction helps to expose small differences in the output of \texttt{layer\_norm} through the amplification of \texttt{linear}. 
Such small differences should also attract attention in testing since they may be magnified by the subsequent APIs and cause significant differences in the final result, thereby causing issues like slow training convergence of models \cite{RealWorldNumericalBugs_2017_di, NumericalBugs_2022_wang}. 

However, traditional API-level testing methods test each API independently and cannot detect such errors via API interactions.
Recently, TitanFuzz \cite{TitanFuzz_ISSTA_2023_Deng} artificially pairs APIs to make up API interactions, but errors found by these artificial interactions are usually considered as not occurring in real model-executing scenarios. For instance, developers regarded issue \#87960 in Fig.~\ref{motivation_not_real_user}, which is caused by an API interaction made up by TitanFuzz, as ``not a real user issue''. 
These suggest the need to consider API interactions in testing, particularly the realistic interactions such that the errors more likely to affect reliability in real model-deploying scenarios can be better exposed. 

\begin{figure}[htbp] 
\centering 

\begin{lstlisting}
import torch.nn.functional as f

output_ln = f.layer_norm(p0, p1, p2, p3, p4)
# Precision Differences: 2.28882e-5

output_linear = f.linear(p0, p1, p2)
# Precision Differences: 0.
\end{lstlisting}

\caption{Example Precision Differences in Testing Single API}
\label{motivation_single_api} 
\end{figure}

\begin{figure}[htbp] 
\centering 
\begin{lstlisting}
import torch.nn.functional as f

output = f.layer_norm(p0, p1, p2, p3, p4)
# Precision Differences: 2.28882e-5
output = f.linear(output, p1, p2)
# Accumulated Precision Differences: 0.17188
\end{lstlisting}
\caption{Example Precision Differences in Testing with API Interactions}
\label{motivation_interaction} 
\end{figure}

\subsection{Frequent Subgraph-Level DL Library Testing}
\label{section_2_2}

\begin{figure}[htbp] 
\centering 
\includegraphics[width=0.74\linewidth]{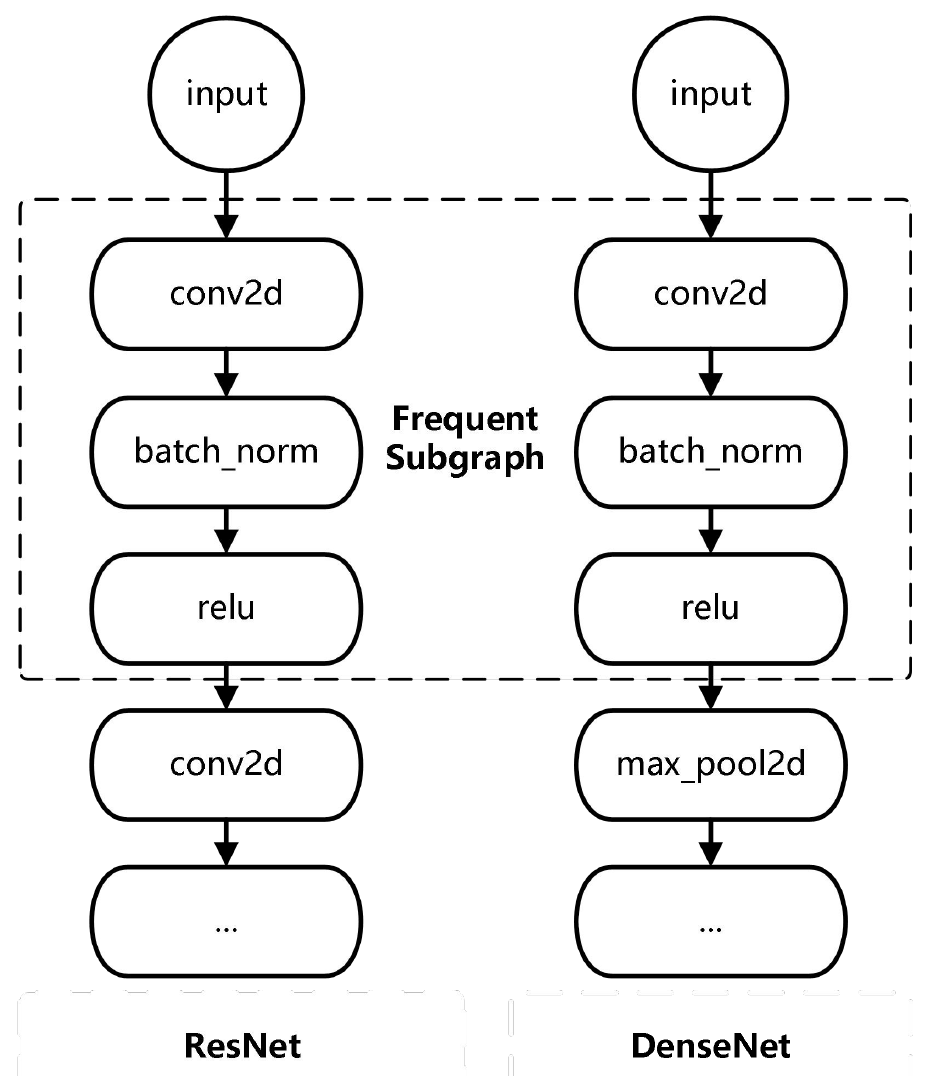} 
\caption{Computation Graphs of ResNet and DenseNet Models}
\label{resnetAdensenetgraph}
\end{figure}

To address the limitations discussed in Section~\ref{section_2_1}, we introduce a new granularity of DL library testing subjects: \textit{frequent subgraphs}. 
Specifically, a deep learning model can be represented as a computational graph, where each node represents an API used in calculation and each edge represents the data flow between APIs. 
Frequent subgraphs refer to the subgraphs frequently appearing in the computation graphs of a series of DL models.
For example, Figs.~\ref{motivation_resnet_code} and ~\ref{motivation_densenet_code} show part of the definition code of ResNet \cite{resnet_2016_he} and DenseNet \cite{densenet_2017_huang} models, respectively. 
Accordingly, we can obtain their computation graphs as shown in Fig.~\ref{resnetAdensenetgraph}.
Then, we can identify frequent subgraphs with a frequency of two from the two models as highlighted with the frame, where the output of \texttt{conv2d} is input to \texttt{batch\_norm}, and the output from \texttt{batch\_norm} is subsequently input to \texttt{relu}. Such frequent subgraphs represent popular real interaction patterns of DL library APIs.

\begin{figure}[htbp] 
\centering 
\begin{lstlisting}
self.features = nn.Sequential(
    nn.Conv2d(3, 32, 3, 1, 1, bias=false),
    nn.BatchNorm2d(32),
    nn.ReLU(inplace=True),
    nn.Conv2d(32, 32, 3, 1, 1, bias=false),
    nn.BatchNorm2d(32)
)
\end{lstlisting}
\caption{Part of the Code Defining the ResNet Model}
\label{motivation_resnet_code} 
\end{figure}

\begin{figure}[htbp] 
\centering 
\begin{lstlisting}
self.features = nn.Sequential(
    nn.Conv2d(3, 64, 7, 2, 3, bias=false),
    nn.BatchNorm2d(64),
    nn.ReLU(inplace=True),
    nn.MaxPool2d(3, 2, 1)
)
\end{lstlisting}
\caption{Part of the Code Defining the DenseNet Model}
\label{motivation_densenet_code} 
\end{figure}

Testing DL libraries at the frequent subgraphs level manifests benefits. 
Specifically, different from the individual APIs in traditional API-level testing, frequent subgraphs introduce API interactions to magnify the small differences likely significantly impacting the calculation result. 
Moreover, the frequent subgraphs identified from real DL models introduce meaningful API interactions rather than artificially made-up combinations whose significance is uncertain. 
As a reminder, using whole DL models to test DL libraries also introduces realistic API interactions, but DL models often include many APIs and are thereby hard to localize the problematic API \cite{FreeFuzz_ICSE_2022_Wei}. Meanwhile, frequent subgraphs are smaller than the whole DL model, while they include common API interactions in popular DL models, which allows prioritizing the validation of these widely-used interactions.
According to the data we collected, a DL model has an average of 339 APIs, with a minimum of 29 APIs and a maximum of 1,143 APIs; meanwhile, subgraphs have only six APIs on average, with a minimum of two APIs and a maximum of seven APIs.
Consequently, testing frequent subgraphs maintains a better efficiency of bug localization.

Besides, we also aim at a new method to prepare meaningful inputs for frequent subgraphs.
Specifically, we record characteristics of the runtime inputs for each API via code instrumentation and generate inputs for the APIs in subgraphs based on the recorded features. 
Unlike DocTer and ACETest relying on the constraints extracted from static documents and codes and thereby are restricted by the constraints able to extract \cite{ACETest_2023_Shi}, we collect the features of real runtime data and generate new inputs following their features. This solution automatically simulates the real inputs without relying on additional information other than the massive and easy-to-obtain real models and datasets. 
Besides, other instrumentation methods often overlook value ranges \cite{FreeFuzz_ICSE_2022_Wei, DeepREL_FSE_2022_Deng, TitanFuzz_ISSTA_2023_Deng}, causing a focus loss of values that may exist in real execution. Different from them, we collect the value range of real inputs and generate inputs residing in that range. We expect these inputs can better simulate the values likely to occur in real model execution, such that we can mine the unaware faults that can be triggered by common inputs, which tend to be more important.

In summary, to address the three limitations mentioned in Section~\ref{section_2_1}, we propose to use frequent subgraphs as the test subjects and generate inputs based on the features of real data inputs by instrumentation. Testing frequent subgraphs helps to focus on testing real API interactions; while instrumentation-based input generation helps to prepare more valid inputs to validate the calculation function of APIs.
Together, these strategies are expected to reveal the bugs missed by the earlier methods considering none or less realistic API interactions and producing many invalid or uncommon inputs.

\section{Methodology}

\begin{figure*}[htb] 
\centering 
\includegraphics[width=0.88\linewidth]{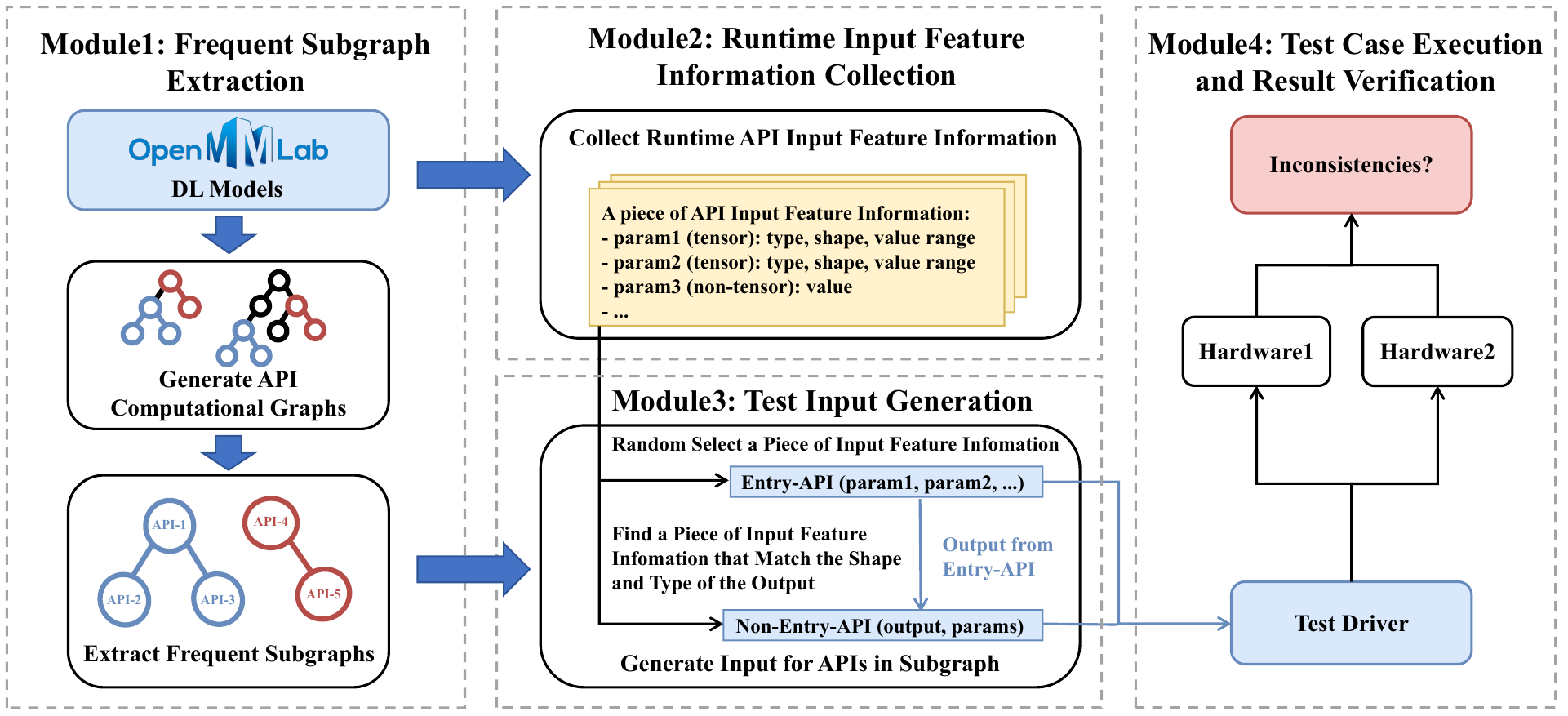} 
\caption{Overview of SORT}
\label{overview} 
\end{figure*}

In this work, we propose a DL library testing method named SORT. SORT tackles the limitations of existing methods by differential testing frequent subgraphs' behaviors with test inputs generated based on real-life inputs' features as elaborated in Section~\ref{section_2_2}. 
As shown in Fig.~\ref{overview}, SORT includes four modules. 
The first module extracts the frequent subgraphs from open-source DL models as the test subjects; 
the second module collects runtime input data features of open-source DL models; 
the third module generates test inputs for frequent subgraphs by referring to the runtime input data;
the last module performs differential testing over frequent subgraphs with the generated inputs across hardware platforms to expose problematic different behaviors, e.g., precision errors.
We elaborate on these modules in the following subsections.

\subsection{Frequent Subgraph Extraction}
\label{section_3_1}

In this section, we introduce our method to prepare frequent graphs that reflect the popular API usage patterns as the test subjects. We first build the computation graphs for popular DL models and then employ a frequent subgraph extraction algorithm to realize the goal.

Specifically, to obtain the computation graph of each DL model, we first collect popular open-source DL models from model repositories like OpenMMLab\footnote{\url{https://openmmlab.com}}. We then analyze the source code of each model to construct the corresponding computation graph, where each API invocation is represented as a node and each data flow is represented as an edge. 
To construct these computation graphs, we iterate through the model code, identify each API function call, and map it to a graph node. Edges are formed based on function signature analysis, where each node’s input-output relationship is tracked and stored in a structured format. 
For example, as shown in Fig.~\ref{resnetAdensenetgraph}, the output of \texttt{conv2d} is used as input for \texttt{batch\_norm} in ResNet and DenseNet; thus this connection is represented by a directed edge between their corresponding nodes in the computation graphs of both models.

Once the complete computation graph for each collected model is built, we apply a frequent subgraph extraction algorithm, gSpan \cite{gspan_2022_yan}, to identify the frequent subgraphs.
gSpan is one of the most popular solutions for mining frequent subgraphs \cite{fsmsurvey_2013_jiang}. It iterates through a set of graphs and conducts a depth-first search to discover and expand subgraphs that meet a minimum frequency threshold, representing the minimum times a subgraph must appear across the model graphs to be considered ``frequent''. 
For example, if we use a frequency threshold of 2, then we can identify a frequent subgraph made up of APIs \texttt{conv2d}, \texttt{batch\_norm}, and \texttt{relu} based on the ResNet and DeseNet models as shown in Fig.~\ref{resnetAdensenetgraph}.

\subsection{Runtime Input Feature Information Collection}
\label{section_3_2}

In this subsection, we introduce how we record the features of runtime data of DL library APIs. Such information is later used to guide the generation of test inputs for frequent subgraphs, which will be introduced in Section~\ref{section_3_3}.

To collect the features of API runtime data, we instrument the source code of each DL library API and run real DL models with real datasets.
Specifically, we hook each API call to record input feature information during each execution. For the tensor parameters, we record feature information about the type, shape, and value range. For the non-tensor parameters, the values are recorded. 
When an API is invoked once in a model with a real input for the model, we will collect a piece of input feature information. 
For example, Fig.~\ref{listing_inputinfoeg} shows a piece of input feature information we collect for API \texttt{conv2d}. It suggests that for the parameter \texttt{input} of API \texttt{conv2d}, we can generate a tensor with a type of float32 and a shape of [16, 3, 224, 224] and values ranging from -102.9072 to 152.3829 as a reasonable test input; for the parameter \texttt{stride}, we can generate an integer test input of 1.  
In this way, we can collect multiple pieces of input feature information for each API after running several models invoking the API with diverse model inputs. The collected feature information guides the later generation of test inputs for the corresponding API.
\begin{figure}[h!] 
\centering 
\begin{lstlisting}[language=java, keywordstyle=\color{black}]
Tensor Parameters:
- input: 
    - type = float32
    - shape = [16, 3, 224, 224]
    - value range = [-102.91, 152.38]
- weight:
    - type = float32
    - shape = [64, 3, 7, 7]
    - value range = [-0.20, 0.30] 
- bias:
    - type = float32
    - shape = [64]
    - value range = [-0.10, 0.10]
Non-Tensor Parameters:
- stride: type = int, values = 1
- padding: type = int, values = 3
- dilation: type = int, values = 1
- groups: type = int, values = 1
\end{lstlisting}
\caption{Recorded Input Feature Information of \texttt{conv2d}}
\label{listing_inputinfoeg} 
\end{figure}

\subsection{Test Input Generation}
\label{section_3_3}

With the collected feature information, we generate test inputs for each frequent subgraph. To systematically generate the input values for parameters, we categorize parameters into three types: \textit{parameters of entry API}, \textit{dependent parameters of non-entry API}, and \textit{non-dependent parameters of non-entry API}. In the following, we explain these parameter types and our method to generate values for them.

\textbullet\  \textit{Parameters of entry API} refer to the parameters of the first API in the mined frequent subgraphs, e.g., parameters of \texttt{conv2d} in the subgraph shown in Fig.~\ref{resnetAdensenetgraph}. 
These parameters need to be generated based on the collected features of runtime data. 
For example, based on the information in Fig.~\ref{listing_inputinfoeg}, we can create a tensor for the \texttt{input} parameter with shape [16, 3, 224, 224] and values randomly sampled from the range [-102.91, 152.38].
The inputs for tensor-type parameters \texttt{weight} and \texttt{bias} can also be generated following the shapes and ranges of their runtime inputs. 
Meanwhile, we can assign the value 1 to \texttt{stride} based on the gathered information, as well as prepare inputs for other non-tensor parameters according to their corresponding values. 

\textbullet\ \textit{Dependent parameters of non-entry API}. In the calculation graph, the non-entry APIs take the outputs from the preceding APIs as the input of some parameters. We call such parameters \textit{dependent parameters}. There is no need to generate inputs for such parameters. 
For instance, as shown in Figs.~\ref{resnetAdensenetgraph}, \texttt{batch\_norm} is a non-entry API. It will automatically take the output tensor from the preceding API \texttt{conv2d} as the value for its dependent parameter \texttt{input}.

\textbullet\ \textit{Non-dependent parameters of non-entry API} refer to the other parameters of non-entry APIs whose values are not the output of previous APIs. For such parameters, we need to prepare input values that are compatible with the dependent parameters of the API. 
Specifically, for a non-entry API that includes dependent parameters and non-dependent parameters, we search for the collected input information record where the shape and datatype of all dependent parameters in the record match the actual shape and datatype of dependent parameters. If multiple records satisfy the requirement, we randomly pick one record. Then, we generate input for each non-dependent parameter based on the recorded features of that parameter.
Consider an example, as shown in Fig.~\ref{resnetAdensenetgraph}, where \texttt{batch\_norm} uses the output of \texttt{conv2d} as its \texttt{input} parameter. For \texttt{batch\_norm}, the parameter \texttt{input} is dependent and others are non-dependent. Suppose that in test execution, the actual \texttt{input} takes the output of conv2d, which is a float32 tensor in the shape of [16, 64, 224, 224]. We then search for the recorded runtime input information of \texttt{batch\_norm} where the \texttt{input} parameter has the same type and shape, i.e., float32 and [16, 64, 224, 224]. We randomly pick a piece of such information, as shown in Fig.~\ref{listing_inputinfoeg_batch_norm}.
Based on this record, we create a tensor for non-dependent parameter \texttt{running\_mean} with shape [64] and values randomly sampled from the range [-0.03, 0.10], and assign the value False to the non-tensor non-dependent parameter \texttt{training}. We prepare inputs for other non-dependent tensor and non-tensor parameters similarly.

\begin{figure}[h!] 
\centering 
\begin{lstlisting}[language=java, keywordstyle=\color{black}]
Tensor Parameters:
- input:
    - type = float32
    - shape = [16, 64, 224, 224]
    - value range = [-1.57, 1.67]
- running_mean:
    - type = float32
    - shape = [64]
    - value range = [-0.03, 0.10]
...
Non-Tensor Parameters:
- training: type = bool, values = False
...

\end{lstlisting}
\caption{Recorded Input Feature Information of \texttt{batch\_norm}}
\label{listing_inputinfoeg_batch_norm} 
\end{figure}

\subsection{Test Case Execution and Result Verification}
\label{section_3_4}

The last step of SORT is to perform differential testing over each frequent graph using the generated test inputs.
We follow existing works \cite{Predoo_ISSTA_2021_Zhang, FreeFuzz_ICSE_2022_Wei, TitanFuzz_ISSTA_2023_Deng} to run the test cases on two different hardware platforms, e.g. CPU and GPU, and compare their outputs. 
If the difference between the outputs obtained on different platforms is within an acceptable tolerance threshold, we consider there is no error; otherwise, potential bugs are reported. 
This criterion is originally suggested by the PyTorch community\footnote{https://discuss.pytorch.org/t/precision-difference-between-gpu-and-cpu/26969,https://discuss.pytorch.org/t/computation-precision-differs-a-lot-between-gpu-mode-and-cpu-mode/23112}, where users introduce the precision issues due to the inherent differences between CPU and GPU and developers highlight the importance of exposing them.

After observing the potential amplification error, we record the output of each API within the subgraph, expecting that the detailed output records may help the developer trace back to the root cause. 
This is inspired by the RIP (Reached, Infected, and Propagated) model in software testing -- when an error is detected on the output of an API, the root cause of the bug should reside in this API and its predecessor APIs.
For example, if we observe differences over \texttt{batch\_norm} in the frequent subgraph shown in Fig.~\ref{resnetAdensenetgraph}, we should check both \texttt{batch\_norm} and \texttt{conv2d} since the problem may be caused by either an issue in \texttt{batch\_norm}, or a bug of \texttt{conv2d} causing a small difference below the threshold but later leading to a non-negligible difference in the final result, or both.

\section{Experimental Setup}

\subsection{Research Questions}
\label{section_4_1}

The experiments in this work aim to address the following research questions (RQs):

\textit{\textbf{RQ1: Can SORT generate more valid input compared to existing API-level testing techniques?}} 
In this RQ, we compare the ratio of valid test inputs generated by SORT to the baseline methods to study SORT's strengths in generating valid inputs.

\textit{\textbf{RQ2: Can SORT effectively identify precision bugs compared to existing API-level testing techniques?}} 
In this RQ, we investigate the number of different types of bugs revealed by the tests generated by different methods to validate SORT's ability to reveal precision bugs.

\textit{\textbf{RQ3: Can API interactions contribute to detecting more bugs?}} 
In this RQ, we compare the effect of testing with and without considering API interaction to understand if the API interaction patterns reflected by frequent subgraphs contribute to the bug-detection ability of SORT.

\subsection{Test Subjects}

In this work, we take PyTorch DL library~\cite{Pytorch_2019_Paszke} as our system under test due to its popularity and widespread use in deep learning applications. 

\subsection{Baselines}
\label{section_4_2}

We compared SORT with existing approaches to understand its effectiveness.
For RQ1 and RQ2, we compared the ratio of valid inputs generated and the types of bugs revealed by FreeFuzz, DeepREL, DocTer, ACETest, and TitanFuzz. 

\begin{itemize}
    \item \textbf{FreeFuzz \cite{FreeFuzz_ICSE_2022_Wei}} collects API runtime input information of tensor data types and non-tensor values from open-source code. Then, it mutates the data type of collected tensor parameters and values of collected non-tensor parameters to prepare new test inputs to conduct differential testing.
    \item \textbf{DeepREL \cite{DeepREL_FSE_2022_Deng}} automatically infers relational APIs based on API documents. It prepares inputs following FreeFuzz for the identified relational API pairs and tests these pairs against the value or status equivalence.
    \item \textbf{DocTer \cite{DocTer_ISSTA_2022_Xie}} extracts input constraints from API documents and generates test inputs satisfying those constraints for fuzzing.
    \item \textbf{ACETest \cite{ACETest_2023_Shi}} calculates the constraints of valid inputs based on the input validation code and generates test inputs satisfying those constraints for fuzzing.
    \item \textbf{TitanFuzz \cite{TitanFuzz_ISSTA_2023_Deng}} leverages large language models to generate and mutate the value of parameters for fuzzing.
\end{itemize}

\noindent For RQ3, to analyze the impact of considering API interactions in bug detection, we introduce an ablation version of SORT.

\begin{itemize}
    \item \textbf{Single-API} is an ablation version of SORT that generates test inputs based on the runtime data but tests each API individually, i.e., creating subgraphs made up of only one API. This ablation version cannot magnify the minor difference of an API using the later APIs.
\end{itemize}

\subsection{Metrics}
\label{section_4_3}

\textbf{Ratio of Valid Inputs \( R \)} refers to the ratio of test inputs that successfully pass all input validity checks to the total number of generated test inputs \cite{DocTer_ISSTA_2022_Xie, ACETest_2023_Shi}, defined as:

\[ R = \frac{{\text{{Number of API inputs that passed validity check}}}}{{\text{{Number of all API inputs}}}} \times 100\% \]

\textbf{Categories of Bug Types.} We categorized the revealed bugs based on the symptoms \cite{TenFuzz_TOSEM_2023_Chen, PytorchEmpirical_2023_ho, symptoms_2021_jia,RealWorldNumericalBugs_2017_di}. Specifically, we consider four categories: \textit{crash bugs}, \textit{NaN bugs}, \textit{precision bugs}, and \textit{performance {degradation} bugs}.
\textit{Crash bugs} refer to instances where the program exits unexpectedly in execution with an error. 
\textit{NaN bugs} and \textit{precision bugs} occur when the API returns a result of NaN or other value that is inconsistent with the expected one, respectively.
\textit{Performance degradation bugs} occur when the program exhibits poor performance or overuses computational resources, such as taking an excessive amount of time to produce results \cite{PytorchEmpirical_2023_ho}. 

\subsection{Implementation}
\label{section_4_4}

\textbf{Extraction of Frequent Subgraphs.}
We collected models implemented in PyTorch and can be successfully run on both CPU and GPU from OpenMMLab model zoo, which results in 49 DL models. 
According to our preliminary experiments, when we set the threshold to be 5, we can collect the top 20\% frequently appearing subgraphs (API interaction patterns). In this work, we use these frequent subgraphs (728 in total) as our test subjects. Considering the popularity of these API interactions, they are expected to have been extensively tested. We try to reveal the bugs in these APIs missed by the existing methods.
There are 15 APIs in these frequent subgraphs: \texttt{flatten}, \texttt{ \_\_mul\_\_}, \texttt{div}, \texttt{softmax}, \texttt{adaptive\_avg\_pool2d}, \texttt{matmul}, \texttt{max\_pool2d}, \texttt{batch\_norm}, \texttt{dropout}, \texttt{relu}, \texttt{conv2d}, \texttt{gelu}, \texttt{linear}, \texttt{layer\_norm}, and \texttt{\_\_add\_\_}. Our evaluation will focus on testing these essential APIs in popular API usage patterns.
As a reminder, \textbf{our approach is not restricted to these 15 APIs}. This setup evaluates whether our method can help reveal the errors that reside in very popular APIs yet are missed by existing methods. By using a lower frequency threshold, SORT can test a broader range of APIs.

\textbf{Collection of Runtime Input Features.} To collect runtime inputs for each API, we executed the 49 popular DL models from OpenMMLab using the datasets recommended by OpenMMLab for each model. For example, we ran the Conformer \cite{conformer_2021_peng} model on the ImageNet \cite{imagenet_2015_russakovsky} dataset and the ResNet \cite{resnet_2016_he} model on the CIFAR \cite{cifar_2009_krizhevsky} dataset to collect runtime input features for the APIs invoked when running the two models.

\textbf{Test Execution.} In our experiments, we ran the extracted frequent subgraphs on both CPU and GPU and required the outputs of these APIs in subgraphs to adhere that any element of the output tensor produced by these executions should satisfy the following condition:
\begin{equation*} |a-b| \leq 10^{-3}\label{eq} \end{equation*}
where \textit{a} and \textit{b} refer to the float-point values at each index in the output tensors in the CPU and GPU executions, respectively.
By setting this criterion following \cite{FreeFuzz_ICSE_2022_Wei, OracleStudy_2019_nejadgholi}, we expect to identify potential precision bugs.

\section{Result Analysis}

\subsection{RQ1: Validity of Input}
\label{section_5_1}

\begin{table*}
    \centering
    \caption{Ratio of Valid Inputs of SORT and Baseline Methods}  
\label{table_ratio_of_valid_inputs} 
\setlength{\abovecaptionskip}{0.2cm}
\setlength{\belowcaptionskip}{0.2cm}
\renewcommand{\arraystretch}{0.20}

\begin{tabular}{cccccccccc}
\toprule
\textbf{API} &  \textbf{FreeFuzz-C} & \textbf{FreeFuzz-P} & \textbf{DeepREL-S} & \textbf{DeepREL-V} & \textbf{DocTer} & \textbf{ACETest-G} & \textbf{ACETest-C} & \textbf{TitanFuzz} & \textbf{SORT}\\
\midrule
flatten
&  75.64\%
& 75.64\%
& 84.75\%
&86.66\%
& 62.23\% 
& 0.10\%
& 0.14\%
& 69.59\%
&100.00\%
\\
\midrule
 \_\_mul\_\_
&   80.26\%
& 80.19\%
& 89.05\%
&
88.97\%
& 28.82\% 
& 44.13\%
& 43.50\%
& 47.42\%
&
100.00\%
\\
 \midrule
 div
&   80.39\%
& 50.43\%
& 88.65\%
&89.93\%
& 71.17\% 
& 44.17\%
& 44.52\%
& 67.02\%
&100.00\%
\\
 \midrule
 softmax
&   54.35\%
& 2.85\%
& 73.86\%
&
/& 33.04\% 
& 22.37\%
& 18.05\%
& 46.88\%
&
100.00\%
\\
 \midrule
 adaptive\_avg\_pool2d&   53.30\%
& 31.05\%
& /&/& / 
& /
& /
& 39.60\%&

100.00\%
\\
 \midrule
 matmul
&   58.79\%
& 4.33\%
& 79.49\%
&
78.13\%
& 2.44\% 
& 6.70\%
& 6.71\%
& 40.97\%
&
100.00\%
\\
 \midrule
 max\_pool2d&   14.38\%
& 0.03\%
& 56.09\%
&/& 3.71\% 
& 0.00\%
& 0.00\%
& 45.58\%
&100.00\%
\\
 \midrule
 batch\_norm&   45.75\%
& 2.01\%
& /&/& 2.44\% 
& 0.15\%
& 0.13\%
& 36.87\%
&

100.00\%
\\
 \midrule
 dropout
&   55.59\%
& 32.86\%
& /&/& 53.62\% 
& 79.74\%
& 79.74\%
& 56.21\%
&

100.00\%
\\
 \midrule
 relu
&   97.66\%
& 3.16\%
& /&
97.94\%
& 74.85\% 
& /
& /
& 53.76\%
&

100.00\%
\\
  \midrule
 conv2d
&   12.04\%
& 0.05\%
& /&/& 0.22\% 
& 0.00\%
& 0.00\%
& 18.86\%
&

100.00\%
\\
 \midrule
 gelu
&   93.72\%
& 0.36\%
& /&/& / 
& /
& /
& 53.91\%&

100.00\%
\\
 \midrule
 linear
&   50.66\%
& 0.30\%
& /&/& 1.29\% 
&  0.15\%
& 0.17\%
& /&

100.00\%
\\
 \midrule
 layer\_norm&   31.67\%
& 0.00\%
& /&
/& 3.86\% 
& 0.00\%
& 0.00\%
& 24.53\%
&

100.00\%
\\
 \midrule
 \_\_add\_\_&   71.38\%
& 71.23\%
& 80.39\%
&
81.83\%
& 27.64\% 
& 27.22\%
& 27.36\%
& 
55.93\%
&100.00\%
\\
\midrule
 \textbf{Total}& \textbf{40.44\%}& \textbf{7.15\%}& \textbf{77.18\%}& \textbf{86.79\%}& \textbf{11.00\%} & \textbf{6.80\%}& \textbf{6.72\%}& \textbf{41.54\%}& 
\textbf{100.00\%}\\
 \bottomrule
\end{tabular}

\end{table*}

To understand the effectiveness of SORT in generating valid inputs, we compared the ratio of valid inputs generated by different methods for all 15 target APIs listed in Section~\ref{section_4_4}.
For a fair comparison, we run each method to generate a total of 15,000 valid test inputs across 15 target APIs. For baseline methods, we tried to generate 1,000 valid test inputs for each API by following their setup and we recorded the total number of test inputs generated\footnote{In cases where generating 1,000 valid test cases required more than 10,000 attempts, we halted further testing for cost efficiency.}. For SORT, we ensure a total of 15,000 valid test inputs for APIs in all subgraphs.

Table~\ref{table_ratio_of_valid_inputs} shows the ratio of valid inputs for each method. Cells in the table marked with a slash indicate that a method cannot generate tests for that API. According to the table, we found that SORT achieved a 100\% pass rate across all target APIs, outperforming the baseline methods. This result demonstrates that the runtime input features during actual model executions provide helpful guidance for generating valid inputs. 

We do case studies to analyze the limitations of each baseline and how SORT tackles the issues. 
Specifically, \textbf{FreeFuzz-C} and \textbf{FreeFuzz-P} achieve a ratio of valid inputs of 40.44\% and 7.15\%, respectively\footnote{FreeFuzz-C uses crashes as a test oracle and FreeFuzz-P uses the correctness of the output as a test oracle.}. FreeFuzz's low ratio of valid inputs can be attributed to its mutation to inputs, which often causes invalid inputs like tensors with an incorrect data type and non-tensors with inappropriate values. 
For instance, FreeFuzz generates an input tensor with a complex data type for the \texttt{relu} function, resulting in a runtime error ``clamp is not supported for complex types'' as shown in Fig.~\ref{freefuzz_invalid_case}. 
In comparison, SORT generates inputs by referring to the features of runtime input data that would not be invalid.
It generates inputs tensor with normal data type float32 for \texttt{relu} and does not trigger runtime errors.
\textbf{DeepREL} uses the input generation approaches of FreeFuzz. The ratio of valid inputs of two DeepREL implementations, DeepREL-S and DeepREL-V\footnote{DeepREL-S uses the status of output as a test oracle and DeepREL-V uses the value of output as a test oracle.}, are 77.18\% and 86.19\%, respectively, which is still not perfect due to similar reasons.
Besides, while DeepREL shows a better ratio of valid inputs than FreeFuzz, it can only test 8 target APIs with equivalent counterparts.

\begin{figure}[htbp] 
\centering 

\begin{lstlisting}[language=python, keywordstyle=\color{black}]
arg_1_tensor = torch.rand([1, 0, 8, 3], dtype=torch.complex128)
arg_1 = arg_1_tensor.clone()
arg_2 = "sum"
res = torch.nn.functional.relu(arg_1,inplace=arg_2,)
# RuntimeError: clamp is not supported for complex types
\end{lstlisting}

\caption{Invalid Test Case Generated by FreeFuzz}
\label{freefuzz_invalid_case} 
\end{figure}

\textbf{DocTer} shows an 11.00\% ratio of valid inputs. Its ineffectiveness results from imprecise input constraints extracted from incomplete documents \cite{DocTer_ISSTA_2022_Xie, ACETest_2023_Shi}.
\textbf{ACETest-C} and \textbf{ACETest-G}\footnote{ACETest-C executes testing on CPU and ACETets-G executes the testing on GPU.} show ratios of valid inputs of 6.80\% and 6.72\%{, respectively}. They generate invalid inputs due to their inability to accurately solve constraints in input validation loops, C++ pointers, and tensor constraints \cite{ACETest_2023_Shi}. When they fail to extract precise constraints, they will randomly generate values, including excessive values, as inputs for APIs. In comparison, SORT avoids this problem by generating input values based on the features of runtime input data, which are more likely to be valid and better simulate real-life inputs.
For example, as shown in Fig.~\ref{acetest_invalid_case}, ACETest generates a test input with an uncommonly large dropout probability, resulting in a runtime error: \textit{``dropout probability has to be between 0 and 1''}. Meanwhile, SORT generates the parameter p as 0.5 for \texttt{dropout} based on the runtime data value and does not trigger runtime errors.

\begin{figure}[htbp] 
\centering 

\begin{lstlisting}[language=python, keywordstyle=\color{black}]
input = torch.rand([1], dtype=torch.float32).cuda()
p = -60442724493381250
train = True
res = torch.dropout(input=input, p=p, train=train,)
# RuntimeError: dropout probability has to be between 0 and 1, but got -6.04427e+16
\end{lstlisting}

\caption{Invalid Test Case Generated by ACETest}
\label{acetest_invalid_case} 
\end{figure}

\textbf{TitanFuzz} achieves a ratio of valid inputs of 41.54\%. The relatively low ratio of valid inputs is due to its uncertainty in generating valid inputs using large language models. As shown in Fig.~\ref{titanfuzz_invalid_case}, TitanFuzz generates an invalid input for \texttt{conv2d}, causing a runtime error related to incompatible shapes: \textit{``shape `[1, 3, 1, 2]' is invalid for input of size 3''}. In comparison, SORT can generate a correct shape for \texttt{conv2d} based on the runtime inputs' features and does not trigger runtime errors, as the example mentioned in Section~\ref{section_3_2}.

\begin{figure}[htbp] 
\centering 

\begin{lstlisting}[language=python, keywordstyle=\color{black}]
input = torch.Tensor([1, 2, 3]).view(1, 3, 1, 2)
kernel = torch.Tensor([[1, 0], [0, 1]]).view(3, 3, 1, 2)
output = torch.nn.functional.conv2d(input, kernel)
# RuntimeError: shape '[1, 3, 1, 2]' is invalid for input of size 3
\end{lstlisting}

\caption{Invalid Test Case Generated by TitanFuzz}
\label{titanfuzz_invalid_case} 
\end{figure}

To summarize, by referring to the extensive features of type, shape, and value range of the runtime inputs for DL library APIs on real datasets, SORT successfully generates more valid test inputs than the baseline methods. Such generated inputs are expected to simulate real execution data of DL libraries and help to concentrate on detecting the errors more likely to happen in real-life usage of DL libraries. In RQ2, we will discuss the bugs revealed with such inputs.

\subsection{RQ2: Effectiveness of Detecting Precision Bugs}
\label{section_5_2}

\begin{table*}[htbp]   
\setlength\tabcolsep{1.4pt}
\renewcommand{\arraystretch}{0.20}
\begin{center}   
\caption{Number of Different Types of Bugs Detected by SORT and Baseline Methods}
\label{table_bug_types} 

\begin{tabular}{c|cccc|cccc|cccc|cccc|cccc}
\toprule
&\multicolumn{4}{c}{\textbf{FreeFuzz}} &\multicolumn{4}{c}{\textbf{DeepREL}}&\multicolumn{4}{c}{\textbf{DocTer}} & \multicolumn{4}{c}{\textbf{TitanFuzz}} &\multicolumn{4}{c}{\textbf{SORT}} \\
& Perf &Crash& NaN& Precision & Perf & Crash& NaN&Precision &  Perf &Crash& NaN& Precision& Perf & Crash& NaN& Precision& Perf & Crash& NaN&Precision\\
\midrule 
flatten
& 0
& 0& 0
& 0
 & 0& 0& 0&0& 0&0
&0
& 0
& 0
& 0
& 0& 0
& 0
& 0
& 0&4
\\
\midrule
 \_\_mul\_\_
& 5
& 0& 0
& 0
 & 0& 2& 0&0& 0&0& 
0& 

0& 0
& 13
& 0& 0
& 0
& 0
& 0&4
\\
 \midrule
 div
& 6
& 0& 0
& 0
 & 0& 3& 0&0& 0&0& 0
& 0
& 0
& 16
& 0& 0
& 0
& 0
& 0&0
\\
 \midrule
 softmax
& 0
& 0& 0
& 0
 & 0& 0& 0&0& 0&0& 
0& 

0& 0
& 7
& 0& 0
& 0
& 0
& 0&0
\\
 \midrule
 adaptive\_avg\_pool2d& 
241
& 0& 0
& 
0
 & /& /& /&/& /&/& /& 
/& 0
& 46
& 0& 0
& 0
& 0
& 0&0
\\
 \midrule
 matmul
& 3
& 0& 0
& 0
 & 0& 0& 0&0& 0&0& 
0& 
0& 0
& 75
& 9& 0
& 0
& 0
& 0&0
\\
 \midrule
 max\_pool2d
& 0
& 0& 0
& 0
 & 0& 0& 0&0& 0&219& 0
& 
0
& 0
& 3
& 0& 0
& 0
& 0
& 0&6
\\
 \midrule
 batch\_norm
& 0
& 0& 0
& 0
 & /& /& /&/& 0&0& 
0& 
0& 0
& 0
& 0& 0
& 0
& 0
& 0&230
\\
 \midrule
 dropout
& 2& 0& 0
& 0
 & /& /& /&/& 0&0& 0
& 
0
& 0
& 
8& 0& 0
& 0
& 0
& 0&152
\\
 \midrule
 relu
& 1
& 0& 0
& 0
 & 0& 0& 0&0& 0&0& 
0& 
0& 0
& 9& 0& 0
& 0
& 0
& 0&86
\\
  \midrule
 conv2d
& 0
& 0& 0
& 0
 & /& /& /&/& 0&17& 0
& 0
& 0
& 14& 0& 0
& 0
& 0
& 0&464
\\
 \midrule
 gelu
& 
0
& 0& 0
& 
0
 & /& /& /&/& /&/& 
/& 

/& 0
& 7& 21& 0
& 0
& 0
& 0&376
\\
 \midrule
 linear
& 0
& 0& 0
& 0
 & /& /& /&/& 0&0& 0
& 0
& /& /& /& /& 0
& 0
& 0&1,258
\\
 \midrule
 layer\_norm
& 0
& 0& 0
& 0
 & /& /& /&/& 0&0& 
0& 

0& 0
& 17& 0& 0
& 0
& 0
& 0&0
\\
 \midrule
 \_\_add\_\_& 10& 0& 0
& 0
 & 0& 2& 0&0& 0&0& 0
& 0
& 0
& 
8& 0& 0
& 0
& 0
& 0&796\\
\midrule
 \textbf{Total}& \textbf{267}& 0& 0& 0 & 0& \textbf{7}& 0&0& 0& \textbf{236}& 
0& 
0&0& \textbf{223}& \textbf{30}& 0& 0& 0& 0&\textbf{3,376}\\
\bottomrule
\end{tabular}
 
\end{center}   
\end{table*}

To investigate the effectiveness of SORT in detecting precision bugs, we counted the bugs identified by different methods in each category. As shown in Table~\ref{table_bug_types},
SORT outperforms all the baseline methods in detecting precision bugs, with a total of 3,376 precision bugs reported. These precision bugs were particularly prevalent in APIs such as \texttt{conv2d} (464 cases), \texttt{batch\_norm} (230 cases), \texttt{linear} (1,258 cases), and \texttt{\_\_add\_\_} (796 cases). 
We reported these precision issues to the PyTorch developers\footnote{https://github.com/pytorch/pytorch/issues/133006}. 
The PyTorch developers added tags of \textit{numerical-stability} and \textit{triaged} for the issues we reported. They stated the impact of difference accumulation: \textbf{\textit{This is indeed a major concern for machine learning as a field.}}

Meanwhile, all baseline methods missed these precision bugs. They generally focused on crash and performance degradation bugs.
For instance, FreeFuzz detected 267 performance degradation bugs, primarily in the \texttt{adaptive\_avg\_pool2d} API (241 cases). These bugs were associated with unexpected performance behavior, where the speed of processing float32 values outperformed float16 values. DeepREL identified 7 crash bugs across \texttt{\_\_mul\_\_}, \texttt{div}, and \texttt{\_\_add\_\_}, typically related to input types and boundary cases. 
DocTer uncovered 236 crash bugs, primarily in \texttt{max\_pool2d} (219 cases) and \texttt{conv2d} (17 cases), caused by corner cases. TitanFuzz found 223 crash bugs, notably in \texttt{matmul} (75 cases) and \texttt{adaptive\_avg\_pool2d} (46 cases), as well as 30 NaN bugs, with a concentration in \texttt{gelu} (21 cases) and \texttt{matmul} (9 cases). Finally, ACETest did not detect any bugs during our experiments.
As a reminder, when reproducing FreeFuzz, DeepREL, DocTer, TitanFuzz, and ACETest methods, we observed a decrease in their error-triggering ability compared to the results reported in their papers. This can be due to the newer version of PyTorch (2.3.0) used in our experiment, where many bugs have been fixed.

In summary, the testing results demonstrate that SORT can detect precision bugs, which are also considered a significant issue threatening the reliability of building and deployment of DL models, while the baseline methods barely detect such bugs. As discussed in RQ1, the improved validity and potentially higher similarity to the real-life inputs of the generated inputs contribute to this superiority. Besides, the consideration of API interactions further helps to expose precision errors as well, which we will discuss in RQ3.

\subsection{RQ3: Effectiveness of API Interactions in Bug Detection}
\label{section_5_3}

\begin{figure}[htbp] 
\centering
\centering
	\begin{subfigure}{1.0\linewidth}
		\centering
		\includegraphics[width=1\linewidth]{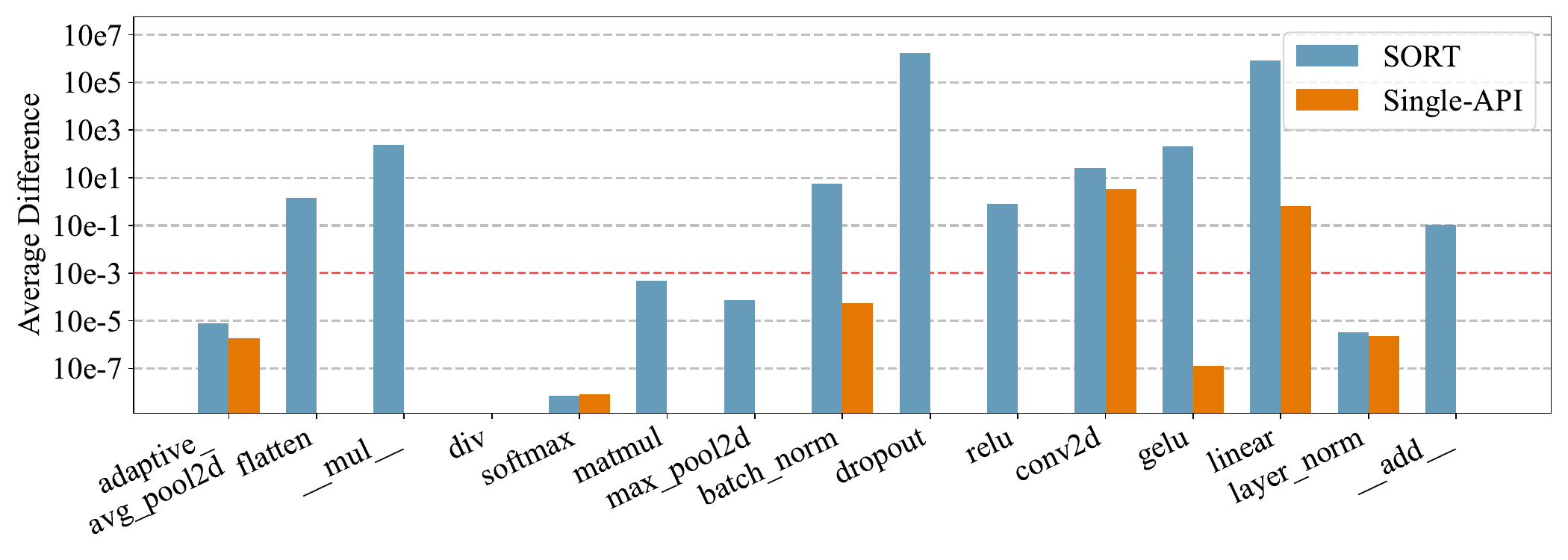}
		\caption{Average Output Difference Observed between CPU and GPU}
		\label{Average_Difference}
	\end{subfigure}
	\begin{subfigure}{1.0\linewidth}
		\centering
		\includegraphics[width=1.0\linewidth]{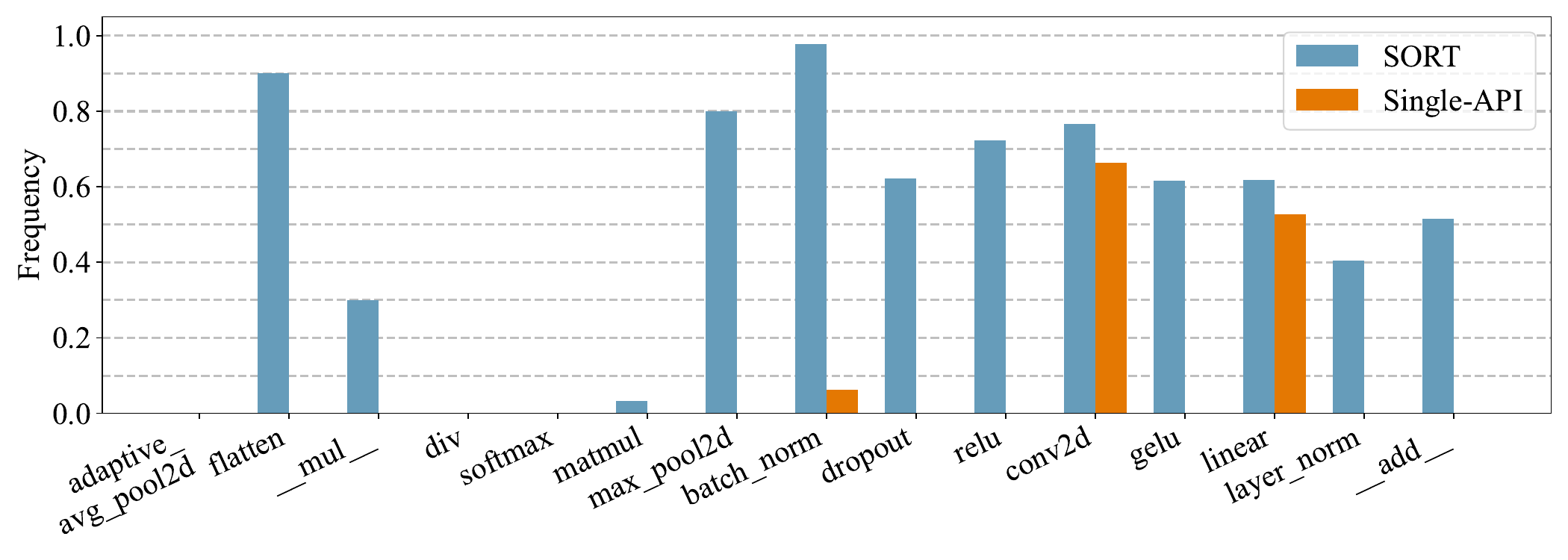}
            \caption{Frequency of APIs Involved in Erroneous Behaviors Revealed}
		\label{Checked_Frequency}
	\end{subfigure}
	
	\caption{Testing Results of SORT and Single-API Testing}
\end{figure}

In this subsection, we investigate whether considering API interactions contributes to the bug-detection ability of SORT. 
First, we compare the average output differences observed between CPU and GPU executions for each API triggered by SORT and single-API testing. 
Fig.~\ref{Average_Difference} shows the comparison, where the red dotted line indicates the threshold to report a bug. Differences below the threshold are considered tolerant precision differences.
We found that \texttt{flatten}, \texttt{\_\_mul\_\_}, \texttt{batch\_norm}, \texttt{dropout}, \texttt{relu}, \texttt{gelu}, and \texttt{\_\_add\_\_} show average observed differences greater than thresholds in SORT but not in single-API testing.
This trend suggests that SORT can magnify the differences between CPU and GPU executions and thereby is more likely to reveal more bugs in APIs.

Next, we further analyze how API interactions help to reveal bugs by magnifying output differences. 
Given that we cannot confirm the buggy APIs, we do not study the explicit relation between API interactions and actual bug detection ability in this work.
Meanwhile, it is well-known that only if an API is involved in an observed abnormal behavior, can the bugs of this API be revealed based on the observed behavior. 
Thus, we study how API interactions help to trigger abnormal behaviors involving more APIs, thereby likely contributing to revealing more buggy APIs.
Specifically, we study the frequency of each API being involved in erroneous behaviors observed in test executions.
This frequency is defined as the ratio of the test executions where an API is involved in erroneous behavior to all executions in which this API is involved. If an API is more frequently involved in observed abnormal behaviors, developers may pay more attention to inspecting this API with the observed behaviors as clues \cite{theoretical_2013_xie}.
As mentioned in Section~\ref{section_3_4}, when SORT reveals a bug, both the API whose outputs violate the difference threshold and its previous APIs in the frequent subgraph relate to the observed abnormal behavior; meanwhile, single-API testing only associates the API under testing with the observed abnormal behavior.

Fig.~\ref{Checked_Frequency} compares the frequency of each API involved in erroneous behaviors revealed by single-API testing and SORT. We found that erroneous behaviors triggered by SORT involve twelve APIs; while single-API testing reveals erroneous behaviors of only three APIs. 
Specifically, nine APIs (i.e., \texttt{flatten}, \texttt{\_\_mul\_\_}, \texttt{matmul}, \texttt{max\_pool2d}, \texttt{dropout}, \texttt{relu}, \texttt{\textbf{gelu}}, \texttt{\textbf{layer\_norm}}, \texttt{\_\_add\_\_}) are never involved in erroneous behaviors in single-API testing since testing these APIs individually never results in an output difference violating the threshold. Meanwhile, SORT triggers erroneous behaviors on the API interactions involving these APIs. 
Besides, in comparison to single-API testing, SORT increased the frequency of being involved in erroneous behaviors of three APIs (i.e., \texttt{\textbf{batch\_norm}}, \texttt{conv2d}, \texttt{linear}), warning a higher priority to check them. 
In particular, the three APIs in bolded texts are never or seldom involved in the revealed erroneous behavior based on the single-API testing results, and as a result, developers may miss the potentially impactful precision loss caused by these APIs. Meanwhile, SORT helps to expose the impact of small output differences in causing non-ignorable value differences to subsequent APIs in popular real-life API interaction usages. 
We reported these cases to the PyTorch developers and they added the \textit{numerical-stability} tag to the corresponding issue reports and already triaged them.

\begin{figure}[htbp] 
\centering 
\includegraphics[width=0.75\linewidth]{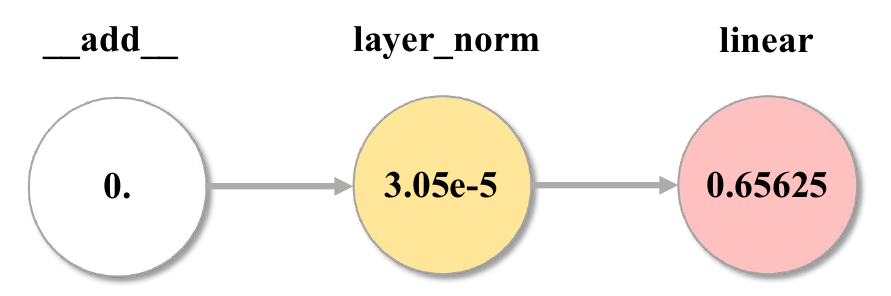} 
\caption{An Example of Subgraph Magnifying Difference}
\label{Checked_Example} 
\end{figure}

Fig.~\ref{Checked_Example} shows an example to illustrate how API interactions contribute to the bug-detection ability of SORT. 
In the figure, each API within the subgraph is represented as a node and the text shows the absolute difference between its outputs on CPU and GPU.
Nodes are painted in red if the difference exceeds the threshold, in yellow if the difference is smaller yet non-zero, and in white if no difference.
In this example, \texttt{linear} shows over-threshold output differences, suggesting that \texttt{\_\_add\_\_}, \texttt{layer\_norm} , and \texttt{linear} are involved in erroneous behaviors and need further checking. Besides, \texttt{\_\_add\_\_} shows a zero output difference, while small differences in an API (e.g., \texttt{layer\_norm}) here accumulate and amplify in subsequent APIs (e.g., \texttt{linear}), leading to above-threshold differences. Thus, \texttt{layer\_norm} will be warned by SORT to check against the potential threat to cause significant precision loss; while it will be ignored in single API testing.

In conclusion, the results demonstrate that considering API interactions contributes to the bug-detection ability of SORT. This capability contributes to SORT’s advantage in detecting precision bugs, as observed in RQ2.

\section{Discussion}
\label{section_6}

To understand whether SORT aligns with developers' expectations of testing methods of deep learning libraries, we conducted a survey with ten developers working with DL libraries. The questions and received feedback of the survey are available in our artifact \cite{artifact}.
Seven of these participants are developers from ICT companies working with DL libraries, and the other three are PhD students with over five years of DL model-building and debugging experience.
The main findings of the survey are as follows:

\textbf{Model-level testing leads to difficulties in error localization and testing API sequences may alleviate this problem.} All participants agreed that it is difficult for developers to trace the root cause (i.e., specific buggy APIs) of errors based on the model-level testing results. They also suggest this hinders the efficiency of debugging and repair practices. 
Meanwhile, 60\% participants believed that bugs identified through testing API sequences of no more than ten APIs are easier to trace and fix. 
As introduced in Section~\ref{section_2_2}, the frequent subgraphs mined by SORT typically include fewer nodes (e.g., 2 to 7 on our subjects of popular DL models), making them easier to debug. Therefore, participants' feedback suggests that frequent subgraph-level testing is likely to offer benefits in alleviating error localization problems, as mentioned in Section~\ref{section_2_2}.

\textbf{Testing commonly-used API sequences tends to reveal more important bugs than testing random combinations.} All participants agreed on the effectiveness of API sequence-based testing in exposing issues found with interaction between APIs. 80\% of them further pointed out that certain API sequences frequently appear in real-world inference or training scenarios and bugs within these frequently used API sequences could affect the performance of many models, making it more meaningful to test commonly-used API sequences rather than random combinations. The feedback highlights the potential benefits of testing commonly used API interaction sequences, confirming the value of focusing on realistic and frequently-used API interaction as mentioned in Section~\ref{section_2_2}.

\textbf{Minor precision differences below the warning threshold may accumulate and finally cause serious issues.} 80\% participants noted that DL library APIs often exhibit precision differences when executed on different hardware platforms and these differences can accumulate over time. They agreed that accumulated differences may eventually lead to issues like non-convergence on certain hardware.
The feedback from participants emphasizes the importance of identifying these minor differences, which reveals the significance of subgraph testing since it can reveal minor differences magnified by API interactions as shown in Section~\ref{section_5_3}.

\section{Threats to Validity}

A threat to validity of our approach is about the completeness of input features recorded based on certain real datasets. Indeed, we cannot guarantee to obtain the complete features of all valid inputs for each API. As a result, we may fail to generate valid inputs for the real usage scenarios not covered by the adopted datasets. However, unlike the limitations of existing methods, such incompleteness does not lead to the generation of invalid test inputs and the miss detection of non-crash bugs. Meanwhile, we have identified a few precision bugs using the information available in our evaluation. Therefore, we consider the evaluation results should demonstrate the effectiveness of our input generation approach. Besides, given the diversity of available DL datasets, supplementing the recorded features with more datasets should not be too challenging.

Another potential threat is about the selection of parameters in frequent subgraph extraction. The choice of the frequency threshold decides the frequent subgraphs available for testing DL library APIs. A too-large threshold may confine testable APIs to only the ones in highly pervasive interaction patterns; meanwhile, prioritizing the tests with patterns at a small frequency (such patterns are likely quite massive) may postpone revealing the bugs more likely to be encountered. In this work, we consider frequent subgraphs present at least five times in the collected models, representing the top 20\% popular API interaction patterns. The users of SORT may follow our practice to set a proper threshold for their test needs.

The last potential threat is about the generalizability of our approach. 
In this work, we evaluate SORT by testing several PyTorch APIs using the popular DL models implemented by OpenMMLab. We use models in OpenMMLab because it is a popular open-source DL model zoo, and we only test PyTorch since OpenMMLab mainly provides PyTorch implementations of models. The idea of SORT is not restricted to these subjects. 
It can also be applied to testing other DL libraries based on the DL models implemented with those libraries from other authentic sources. One of our future works would be applying SORT to reveal bugs on other widely-used DL libraries with other DL models and datasets.

\section{Related Works}

DL library testing is an essential research domain for ensuring the reliability and accuracy of DL applications. Many testing methods have been proposed to realize the goal.

CRADLE \cite{CRADLE_ICSE_2019_Pham} is the first work to detect bugs in DL libraries. It works at the model level, i.e., using whole DL models as test inputs of DL libraries and then localizing the buggy APIs. Later model-level testing methods mainly focus on a diverse mutation of models \cite{LEMON_FSE_2020_Wang, Audee_ASE_2020_Guo, Muffin_ICSE_2022_Gu}. These model-level testing methods have identified a few real bugs, but they introduce complexity in fault localization due to generally massive APIs in a model.

API-level DL library testing methods tackle the shortcomings of bulky model-level testing by testing APIs in a DL library individually. 
In particular, Predoo \cite{Predoo_ISSTA_2021_Zhang} emphasizes the importance of focusing on the precision error of DL APIs and studies the precision errors of seven specific APIs. DocTer \cite{DocTer_ISSTA_2022_Xie}. FreeFuzz \cite{FreeFuzz_ICSE_2022_Wei}, ACETest \cite{ACETest_2023_Shi}, and TitanFuzz \cite{TitanFuzz_ISSTA_2023_Deng} introduce documentation-guided, code-guided, and large language model-driven methods to produce test inputs.

Considering the potential helpfulness of API interactions for bug detection, later methods create API combinations \cite{GraphBased_ICSE_2021_Luo, Muffin_ICSE_2022_Gu, COMET_TOSEM_2023_Li}.
Prior works focus on achieving higher API combination diversity.
They create a whole model as the test subject as the model-level testing methods do and thereby also suffer from difficulties in fault localization. 
Besides, EAGLE \cite{EAGLE_ICSE_2022_Wang} employs equivalence rules for testing. Some of its equivalence rules involve API combinations, but no errors were detected with such combinations. 
In addition, TitanFuzz \cite{TitanFuzz_ISSTA_2023_Deng} uses large language models to generate API combinations, while many reported bugs are only triggered by the specific API sequences artificially made, which are rarely used in real-world DL models.
These methods do not take care of the real API interactions that tend to be more important since they are more likely to be frequently used in daily life and thereby their bugs tend to lead to more severe consequences.

In this paper, we introduce a novel DL library testing method, SORT, which takes the test subjects at a granularity distinct from existing methods, i.e., frequent subgraphs of DL model computation graphs. Frequent subgraph-oriented testing introduces meaningful API interactions while maintaining the efficiency in fault localization, and finally helps to detect more precision bugs. 
Besides, SORT enhances the generation of test inputs. Unlike existing works that rely on constraints extracted from static documents and codes \cite{DocTer_ISSTA_2022_Xie, ACETest_2023_Shi}, SORT generates test inputs by referring to the real runtime inputs features on easily obtained real DL datasets to simulate real valid inputs. Also, unlike existing methods overlooking the value ranges of generated inputs \cite{FreeFuzz_ICSE_2022_Wei,DeepREL_FSE_2022_Deng,TitanFuzz_ISSTA_2023_Deng}, SORT collects extensive input features about the type, shape, and value ranges of real inputs and generates inputs within these ranges to better simulate real executions, expecting to reveal faults likely to happen in real-life usage of DL libraries.

\section{Conclusion}

In this paper, we propose a novel DL library testing method called SORT to address the limitations of existing methods in leveraging representative API interactions and preparing meaningful test inputs to test DL libraries.
Specifically, to reflect meaningful API interactions while maintaining fault localization efficiency, SORT uses a new type of test subjects, i.e., frequent subgraphs of model computation graphs. 
Besides, SORT collects extensive runtime input features of each API under executing real-life data and generates test inputs likely more valid and meaningful 
based on these features. 
Evaluation results demonstrate SORT effectively reveals many unaware precision issues on popular PyTorch APIs, by generating much more (100\%) valid inputs and properly magnifying minor yet impactful output differences via meaningful API interactions. 

\section*{Acknowledgment}
This work was partially supported by National Key R\&D Plan of China (Grant No.\url{2024YFF0908003}), National Natural\\
Science Foundation of China (No.~\url{62472326}), and CCF-Zhipu Large Model Innovation Fund (No.~\url{CCF-Zhipu202408}).

\bibliographystyle{IEEEtran}
\bibliography{FrameworkTesting}

\end{document}